# AI analysis of medical images at scale as a health disparities probe: a feasibility demonstration using chest radiographs

**Authors:** Heather M. Whitney*[1], Hui Li[1], Karen Drukker[1], Elbert Huang[2], Maryellen L. Giger[1]

**Affiliations**
[1]The University of Chicago, Department of Radiology, Chicago, IL, USA
[2]The University of Chicago, Department of Medicine, Chicago, IL, USA

*first author

Corresponding author: Heather Whitney (hwhitney@uchicago.edu),
The University of Chicago
Department of Radiology
MC 2026
5841 South Maryland Avenue
Chicago, Illinois 60637
(773) 702-6778
ORCID 0000-0002-7258-1102

**Objectives:** Health disparities (differences in non-genetic conditions that influence health) can be associated with differences in burden of disease by groups within a population. Social determinants of health (SDOH) are domains such as health care access, dietary access, and economics frequently studied for potential association with health disparities. Evaluating SDOH-related phenotypes using routine medical images as data sources may enhance health disparities research. We developed a pipeline for using quantitative measures automatically extracted from medical images as inputs into health disparities index calculations.

**Methods:** Our study focused on the use case of two SDOH demographic correlates (sex and race) and data extracted from chest radiographs of 1,571 unique patients. The likelihood of severe disease within the lung parenchyma from each image type, measured using an established deep learning model, was merged into a single numerical image-based phenotype for each patient. Patients were then separated into phenogroups by unsupervised clustering of the image-based phenotypes. The 'health rate' for each phenogroup was defined as the median image-based phenotype for each SDOH used as inputs to four imaging-derived health disparities indices (iHDIs): one absolute measure (between-group variance) and three relative measures (index of disparity, Theil index, and mean log deviation).

**Results:** The iHDI measures demonstrated feasible values for each SDOH demographic correlate, showing potential for medical images to serve as a novel probe for health disparities.

**Conclusions:** Large-scale AI analysis of medical images can serve as a probe for a novel data source for health disparities research.

**Advances in knowledge:** We have demonstrated the feasibility of using data extracted from medical images as inputs to health disparities indices, making possible their future use in data dashboards.

**Keywords**: AI, health disparities, medical imaging, SDOH, health disparities

## 1 Introduction

Health disparities (HDs) are differences in the burden of disease experienced by different groups within a population.[1] They have been tied to variation in social conditions over the course of a lifetime also known as social determinants of health (SDOH).[2,3] The United States Department of Health and Human Services has outlined five domains of SDOH, incorporating health care access and quality, built environment, social and community context, economics, and education at both the individual and structural level. The National Health Service in the United Kingdom outlines the relevant domains as protected characteristics, socioeconomic deprivation, social vulnerability, and geography.[6] SDOH variables include economic stability, educational access, social cohesion, neighborhood-level racial segregation, and health care access. Distinctive clusters of these non-medical factors can be considered SDOH phenotypes.[7] There is evidence that in the United States, SDOH accounts for a substantial proportion (47%) of health outcomes at the county level, even more than health behaviors (34%), clinical care (16%), or physical environment (3%).[8] Using SDOH to study population-level health disparities can have a substantial impact for health outcomes, contributing to targeted interventions and advocacy.[9]

Many measures of health for HD studies are currently collected from numerical or text-based information in EHRs, such as laboratory results or natural language processing of provider notes, such as is done for diabetes[10] or patient adherence to medication.[11] Aggregations of these clinical data has been shown to correlate to disease and be studied in the context of SDOH to identify population-level HDs and address them, including studying patient concerns and challenges,[12–14] and tailoring interventions.[15–17] Recent advances have led to genomic data also being used to identify SDOH phenotypes (such as for accelerated aging[18]).

Notably, many electronic health records (EHRs) are limited in their ability to capture full SDOH data for all patients. Ideally, data fields from standardized sources such as the PhenX Toolkit[19,20] would be incorporated into EHRs and collected from all patients, but this is a difficult task to enact across health systems, especially retroactively. Recent work has reported that standard EHR demographic data of race, ethnicity, sex, age, and median annual income in census tracts correlate to dyads, triads, and tetrads of SDOH (food insecurity, social isolation, daily stress, and housing and utilities),[7] resulting in 'SDOH phenotype demographic correlates.' Including the latter broadens the types and amount of data that can be studied by HD researchers. For simplicity, here we group together SDOH phenotypes and SDOH phenotype demographic correlates under the term 'SDOH-related phenotypes.'

Detecting SDOH-related phenotypes using alternative data sources, such as medical images, may also enhance research and development of services to address HDs. 'Big data' computational methods can yield, at scale via high-throughput means, computer-extracted radiomic features of medical images (termed *radiomic phenotypes*) across multiple large groups (i.e., thousands to millions of individuals) due to high-throughput segmentation and feature extraction algorithms. In addition to traditional radiomic features, which are human-engineered mathematical descriptors of medical images or image findings, features directly extracted from images using deep learning networks can also serve as 'surrogate' imaging-based features. In this work, for simplicity, we refer to any feature extraction from a medical image as a 'radiomic feature.' Radiomic features have been extensively studied, leading to robust, generalizable feature extraction methods for multiple purposes.[21–23] Traditional radiomic features describe the appearance of anatomy/physiology of pixels in a region of interest — such as parenchymal texture of the breast, parenchymal measures of the of the lung, or interstitial nodular patterns in the lung — describing either ‘normal’ or diseased tissue. Traditional radiomics features include morphological features, such as aortic knob diameter from chest radiographs, which can be used to evaluate early changes of the aortic structure,[24] and texture features, such as those describing breast parenchymal texture on otherwise ‘normal’ screening mammograms.[25] The biological relevance of radiomic features has been demonstrated in pre-clinical animal studies for their ability to characterize changes in tissue compared to controls, e.g., in lung tissue due to environmental toxin exposure[26] or fibrotic remodeling,[27] and excess deposition of extracellular matrix in the liver.[28] Some small preliminary studies have also observed differences in the image-based appearance of tissue by demographic attributes that correlate with SDOH, e.g.,[29–31], contributing to the new fields of socioradiomics and socioradiogenomics.[32]

Using ‘big data’ of medical imaging to contribute to HD research could establish a completely new paradigm, resulting in its direct inclusion in the toolbox of resources to identify and monitor HDs of populations (**Figure 1**), allowing HD researchers to discover associations between radiomic phenotypes and SDOH-related phenotypes.

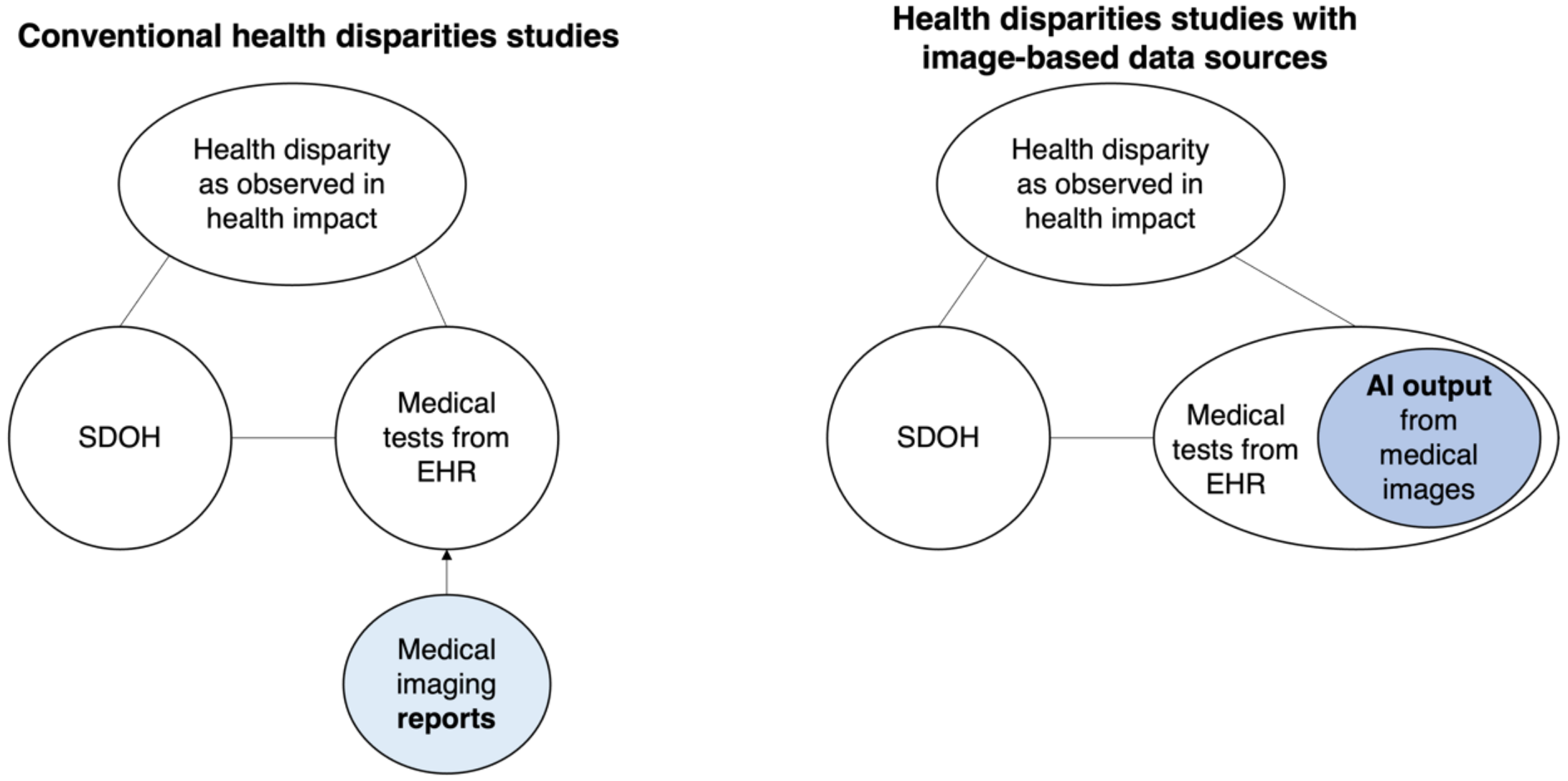

**Figure 1a (left):** Conventional health disparities studies including discovery between SDOH and medical tests to yield health disparity as observed in terms differences of health rates. Note also that medical images often present indicators of differences of health rates, with interpretations (radiology reports) sometimes included in the EHR. However, such radiologic interpretations are (i) manual, (ii) are not performed at scale, (iii) are retained only for the disease/conditions in question, and are not within the EHR at the pixel level.
**Figure 1b (right):** On the other hand, 'big data'-extracted image-based phenotypes from AI analysis can yield, at scale, objective characteristics of both abnormal and normal presentations in medical images, which can be substituted for existing EHR plus additional data in health disparities discovery studies. Thus, we propose in this paper to create a medical imaging-based platform (probe & dashboard) to support health disparities researchers.

Medical images could provide unique data for identifying and monitoring a number of HDs, but this potential is currently untapped. We propose that routine clinical medical images — especially, those deemed negative for a given diagnostic task — can be repurposed as an innovative valuable data source for HD researchers. These images can support ongoing efforts to understand and mitigate HDs, by enabling large scale AI analysis of clinical medical images as an HD probe (**Figure 2**).

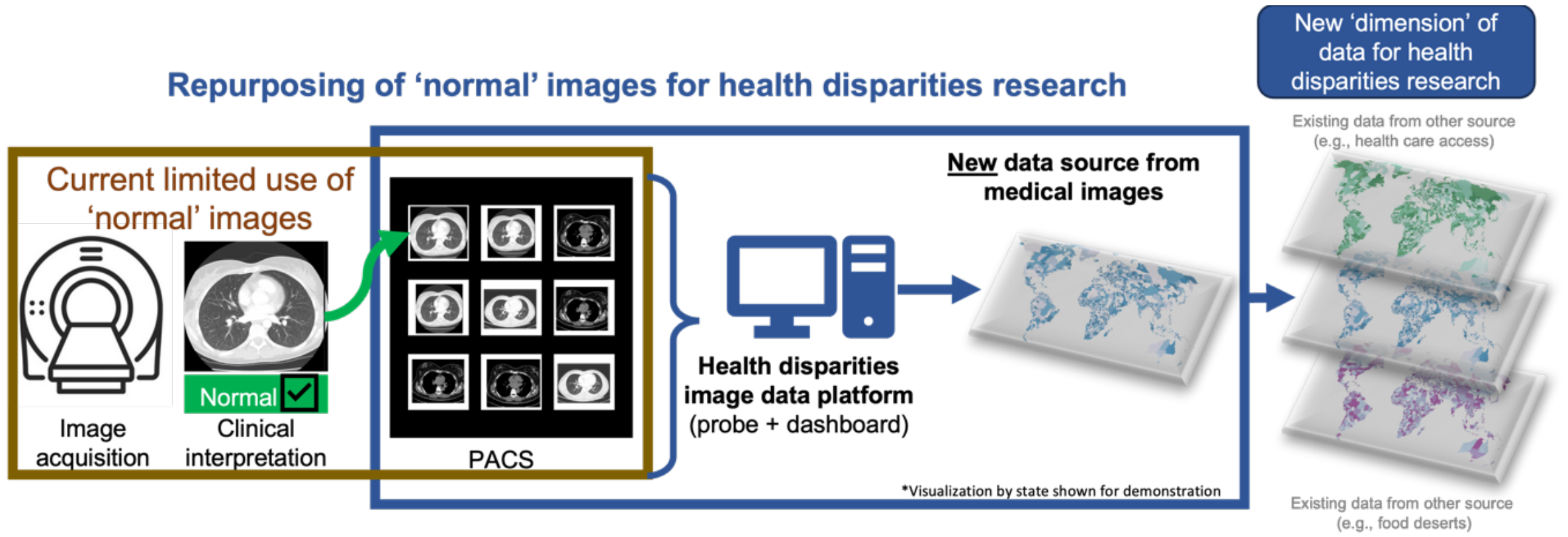

**Figure 2:** Conceptual overview of the medical imaging 'big data' health disparities (HD) platform for (a) the repurposing of 'normal' images for continuous monitoring of HDs and (b) an additional data source for HD research.

In this paper, we report our efforts to create a medical imaging-based platform (probe + dashboard) to support HD researchers. We demonstrate and evaluate the feasibility of our proposed HD probe, using medical images as data sources for HD indices and focusing on the initial use-case of data computer-extracted from chest radiographic images.

## 2 Methods

### *2.1 Dataset*

A retrospective dataset of de-identified and HIPAA-compliant chest radiographs from 1,571patients had been collected under IRB approval from our institution. For each patient there were two types of radiographs: one standard chest radiograph image and one soft tissue radiograph image, which had been calculated using a ClearRead bone suppression algorithm (Riverain Tech). The images had been acquired between January 30, 2020 and February 3, 2021. At the time of imaging, all patients had tested negative for COVID-19 and were deemed 'normal' for the purposes of this study. Demographic information of sex and race was collected from the clinical records and served as SDOH phenotype demographic correlates in this study since SDOH information was not available. For this demonstration study, patients from race groups with less than 20 patients were removed (American Indian or Alaska Native, N = 3; Native Hawaiian/Other Pacific Islander, N = 2). Patients from the groups of ‘More than one race’ (N = 58) and ‘Not reported’ (N = 29) were also removed due to the lack of data available for their SDOH category of race. After these exclusions, our study had a final dataset of 1,571 unique patients (**Table 1**).

**Table 1:** Demographic characteristics (i.e., SDOH demographic correlate subgroups) of the patients in the study. Percentages do not add to 100% due to rounding. Maximum age listed as ‘>89> when the maximum patient age in a subgroup was greater than 89 years old.

| Sex | Race | Number (%) | Age at imaging, years (median, range) |
|---|---|---|---|
| Female | Asian | 8 (<1) | 31 [21, 77] |
| Male | Asian | 16 (1) | 57 [18, >89] |
| Female | Black/African American | 631 (40) | 64 [18, >89] |
| Male | Black/African American | 693 (44) | 42 [20, 81] |
| Female | White | 85 (5) | 55 [18, >89] |
| Male | White | 138 (8) | 63 [19, >89] |

*2.2 Determination of phenogroups*

In this demonstration study, we used a deep learning model that has been previously described[33] to predict the likelihood of pneumonia for the standard and soft tissue chest radiographs for each patient. In this study, these likelihoods were used as surrogate features to describe the radiomic phenotypes of the lung parenchyma in these ‘normal’ patients. First, surrogate features were calculated directly from the two types of chest radiographs (regular and soft-tissue acquisitions) using the deep learning model (Figure 3). Second, the surrogate features derived from each image type were projected via unsupervised principal component analysis (PCA)[34,35] in anticipation of future workflows that might include more than two features (e.g., more than two quantitative values extracted from images). Third, Hotelling’s t-squared statistic[36] was calculated and served as the single numerical output for each patient. This value served as the overall ‘parenchymal score’, i.e., *R*, the radiomic phenotype for each patient. Lastly, unsupervised clustering on *R* was then accomplished using k-means clustering[37] on the study population (i.e., all patients) with the number of clusters optimized by the algorithm and not chosen *a priori*. This was conducted to separate *R* into 'phenogroups', as similarly calculated by other investigators but for non-imaging phenotypes.[38,39]

### *2.3 Calculation of imaging-derived health disparities indices (iHDI)*

The general concept for an iHDI is based upon the established HD*Calc framework,[40] which was designed at the National Cancer Institute (NCI) to provide a collection of measures to evaluate and monitor HDs. The indices in the HD*Calc framework were published in theoretical[41] and case-study format[42] by the NCI in 2005 and 2007, respectively and are comprised of 11 relative and absolute measures of disparity.[43] The HD*Calc framework typically involves inputting EHR health rate data (e.g., percent of the population with 'high' blood pressure and 'high' body mass index) into the health disparities measures.

Calculating an iHDI using imaging-derived data requires the conversion of patient-level data to a health rate by group. Thus, for each of the phenogroups from the unsupervised clustering, we calculated the health status of each SDOH-related group as the ratio of (a) the number of patients in the SDOH-related group with *R* greater than the median *R* to (b) the number of patients in the SDOH-related group (Figure 3).

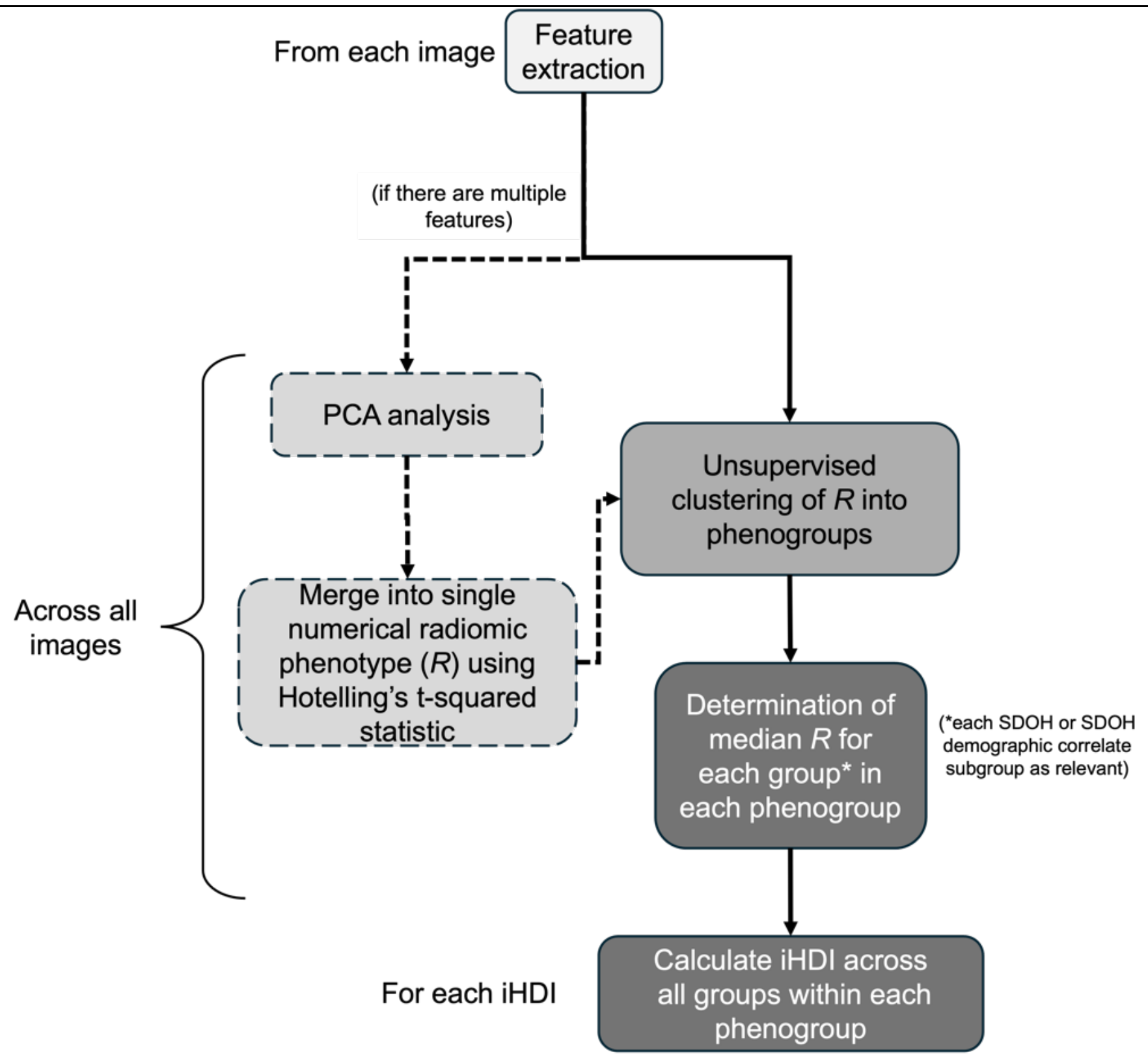

**Figure 3:** Overview of the workflow for calculating an imaging-derived health disparities index (iHDI). SDOH: social determinants of health.

Four forms of measures from the HD*Calc framework can serve as iHDIs to measure differences between groups: (a) *absolute difference* as measured by between-group variance[44] (BGV), (b) *relative difference* as measured by index of disparity[45] ($ID_{isp}$), (c) Theil index[46] (T), and (d) mean log deviation[47] (MLD). These are applicable to imaging-derived health rates due to the unordered nature of the SDOH-related phenotype data that are aggregated. Note that there are other measures available in HD*Calc, but they depend upon ordered data, which is outside of the scope of SDOH-related phenotype data from medical images. Below, we describe the indices and refer the reader to the specific equations and interpretations provided at the HD*Calc Help System website.[48]

*2.3.1 Absolute difference: Between-group variance (BGV)*

The between-group variance[49] (BGV) is an absolute difference measure that squares deviations from a population average and reports them as a sum, weighting by population group size. It reports the variance that would exist if each patient had the median health of their group.

*2.3.2 Relative difference: Index of Disparity ($ID_{isp}$)*

The index of disparity[45] ($ID_{isp}$) is a relative difference measure that (1) sums the absolute differences in a health status between several group rates and a reference rate and (2) reports them as a proportion of the reference rate. It is a modified coefficient of variation measure. In this study, the reference group was set to be the 'White' group for the SDOH demographic correlate of race and the 'Male' group for the SDOH demographic correlate of sex. The authors of HD*Calc recommend using the group with the 'best' health outcome, but because this is undefined for the lung parenchyma score, we decided to use these reference groups because our population is drawn from the United States.

*2.3.3 Relative difference: Thiel Index*

The Thiel Index[50] (T) is a relative difference measure that uses the ratio of the health status of each group to the average health status of the population, weighted by the population proportions, to measure general disproportionality. It has a range of 0 to 1, where 0 represents perfect equality and 1 represents complete inequality.

*2.3.4 Relative difference: Mean Log Deviation (MLD)*

The Mean Log Deviation[51] (MLD) is a relative difference measure that also measures general disproportionality by using a natural logarithm applied to ratio of the health status of each group to the average of the population (i.e., all patients in a dataset). These are summarized via a summation.

### *2.4 Statistical analysis*

We calculated the iHDIs for the lung parenchyma score for each group within each SDOH demographic correlate (sex and race) (see Table 1) and rounded each to two significant digits. Since the goal of this work was to introduce the concept of our novel image-based health disparities probe, we have chosen to present a relatively qualitative demonstration here, with more quantitative analyses to be the focus of future collaborations with HD researchers.

## 3 Results

### *3.1 Phenogroups*

Unsupervised clustering on the surrogate features resulted in two clusters, i.e., phenogroups (Figure 4), with assignment of patients to one of the phenogroups (Table 2, Table 3). Note that the number of clusters was not determined *a priori* but rather through an unsupervised optimization routine within the clustering algorithm.

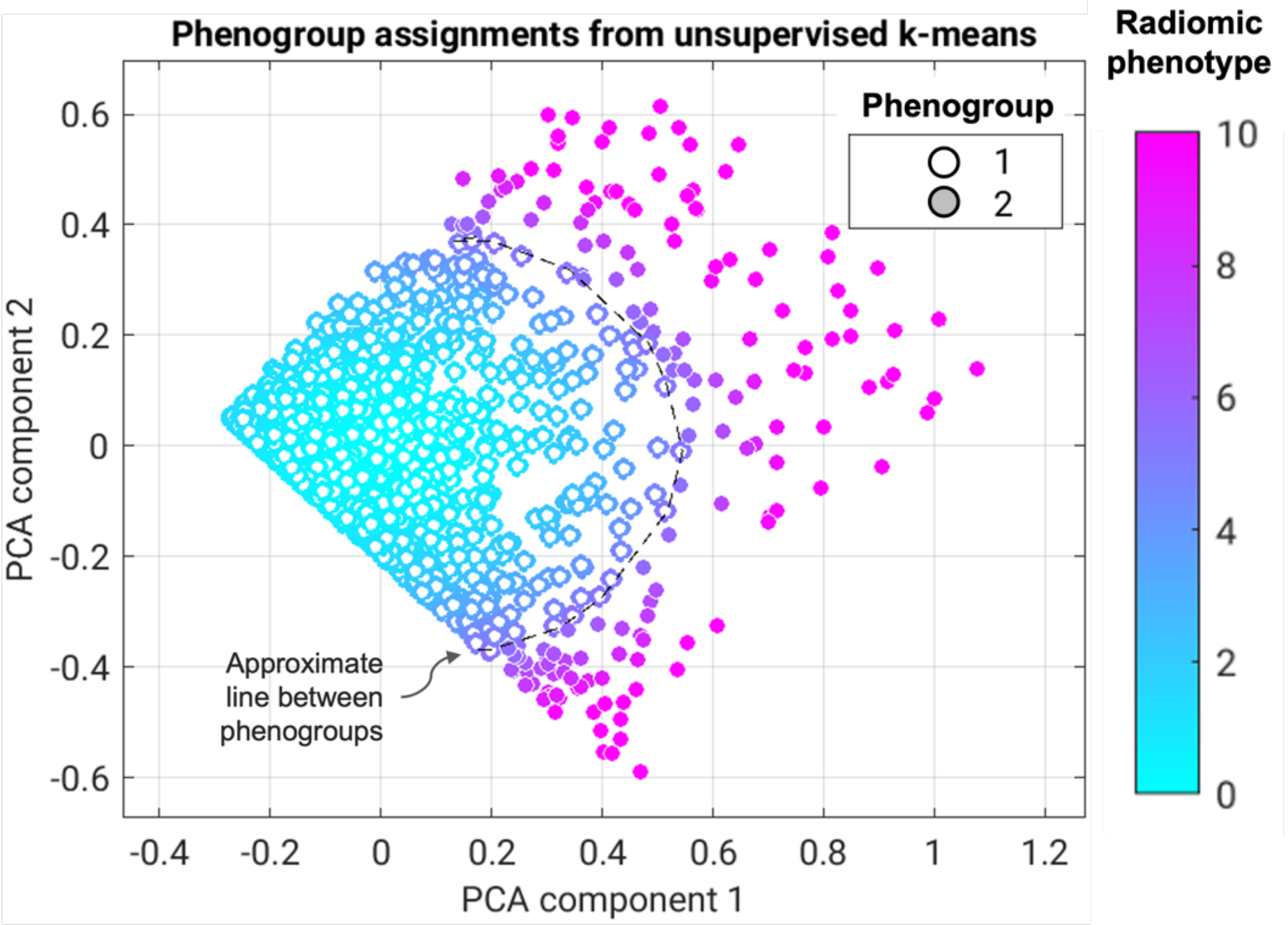

**Figure 4:** Clustering of radiomic phenotype (lung score) from chest radiographs in a dataset of *normal* 1,571 patients. Each datapoint is a patient. The data was clustered into phenogroups by a k-means algorithm on PCA dimension reduction on the lung parenchyma radiomic phenotype *R*.

**Table 2:** Description of patients by the two resulting phenogroups from unsupervised clustering on the radiomic features, reported for the SDOH demographic correlate of race. The median parenchymal score for all patients within each phenogroup is also shown along with the 95% confidence interval (CI).

| Race | Phenogroup | Number of patients | Percentage of phenogroup | Parenchymal score ***R*** (median [95% CI]) |
|---|---|---|---|---|
| Asian | 1 | 21 | 1% | 0.626 [0.043, 3.824] |
| Black/African American | 1 | 1187 | 85% | 0.798 [0.066, 4.524] |
| White | 1 | 203 | 14% | 0.800 [0.102, 3.775] |
| *All* | *1* | *1411* | *100%* | *0.795 [0.068, 4.488]* |
| Asian | 2 | 3 | 2% | 7.478 [6.889, 11.815] |
| Black/African American | 2 | 137 | 86% | 8.998 [5.629, 18.392] |
| White | 2 | 20 | 13% | 8.932 [5.577, 16.819] |
| *All* | *2* | *160* | *100%* | *8.981 [5.62, 18.214]* |

| **Table 3:** Description of patients by the resulting phenogroups from unsupervised clustering on the radiomic features, reported for the SDOH demographic correlate of sex. The median parenchymal score for all patients within each phenogroup is the same as in Table 2 but repeated here for reference. | | | | |
|---|---|---|---|---|
| **Sex** | **Phenogroup** | **Number of patients** | **Percentage of phenogroup** | **Parenchymal score *R* (median [95% CI])** |
| Female | 1 | 636 | 45% | 0.715 [0.060, 4.722] |
| Male | 1 | 775 | 55% | 0.833 [0.087, 4.119] |
| *All* | *1* | *1411* | *100%* | *0.795 [0.068, 4.448]* |
| Female | 2 | 88 | 55% | 8.897 [5.672, 18.045] |
| Male | 2 | 72 | 45% | 9.26 [5.593, 19.697] |
| *All* | *2* | *160* | *100%* | *8.981 [5.62, 18.214]* |

### *3.2 Imaging-derived health disparities indices (iHDI)*

The calculation of the iHDIs for each demographic correlate and each image-based phenogroup demonstrate the potential for imaging-based values to be aggregated into standard health disparities indices (Table 4).

| **Table 4:** Imaging-derived health disparities indices (iHDIs) for each SDOH demographic correlate and each image-based phenogroup, reported as the estimate as defined in HD*Calc.[40] | | | | | |
|---|---|---|---|---|---|
| **SDOH demographic correlate** | **Phenogroup** | **Between group variance (BGV)** | **Index of disparity ($ID_{isp}$)** | **Theil index (T)** | **Mean log deviation (MLD)** |
| Race | 1 | 0.017 | -119% | 0.092 | 0.173 |
| | 2 | 0.019 | -147% | 0.097 | 0.200 |
| Sex | 1 | 0.0018 | -140% | 0.014 | 0.015 |
| | 2 | 0.00016 | -195% | 0.0012 | 0.0012 |

For the iHDI of BGV, BGV tended to be higher for the SDOH demographic correlate of race in both phenogroups than for sex. This suggests the demographic groups of race have larger differences between them in terms of the radiomic phenotype than do the demographic groups of sex. Conversely, $ID_{isp}$ for the SDOH demographic correlate of sex (measuring difference from the Male group in this study) was negative and large, suggesting that the radiomic phenotype for women was 40% and almost 100% less than for men for phenogroups 1 and 2, respectively, while $ID_{isp}$ for the SDOH demographic correlate of race was approximately 19 and 47% less for the non-White groups than for White patients. The Theil Index T suggested that the radiomic phenotype show group similarity within the SDOH demographic correlate of race (0.092 and 0.097 for phenogroup 1 and 2, respectively). However, T was an order of magnitude different between the phenogroups for the demographic correlate of sex (0.014 in phenogroup 1 and 0.0012 in phenogroup 2), suggesting that there were more inter-group differences in radiomic phenotype in phenogroup 2. As expected, the MLD results showed the same trend as the results for T, as both measures are population-weighted and utilize natural logarithm of health rates.

## 4 Discussion

We believe this is the first reported analysis of using medical image-based features as inputs to an HD index. This positions medical images to serve as a novel probe that can be utilized in HD research and importantly, can lead to re-use of medical images that would otherwise sit in storage without further benefit to society. Notably, while most data used for HD research comes from medical records such as routine laboratory tests and clinical information, automatic extraction and aggregation of EHR data can be difficult and time consuming due to differing data format standards. On the other hand, medical images are acquired in a readily-accessible (DICOM) format,[52–54] making routine computerized analyses straightforward once a pipeline is set up. While this study focused on surrogate features from the output of a DL model on chest radiographs, the methods can easily be applied to human-engineered radiomic features and medical images from any imaging modality.

Dashboards for measuring HDs are important tools that aggregate data and analysis into platforms that support population-level research and inform the development of mitigation strategies.[55] Creation of a clinical medical imaging dashboard for HDs could accelerate detection and monitoring of HDs beyond current capacity and timescales, enhance the design and implementation of interventions, and lead to improved health outcomes. Continuous monitoring of HD data is especially crucial. However, a recent review from 2023 noted that there are less than 20 fully developed dashboards as well as substantial limits to their technological design and the number of SDOH-related variables they cover. Additionally, to date there are no dashboards of this type that incorporate AI analyses from medical images as a data source. The development of pipelines for creating such dashboard could yield HD metrics through routine health care allowing for more real-time assessment of drivers rather than relying solely on periodic local (e.g., the Healthy Chicago Survey[56]) or national epidemiologic (NHANES[57]) studies. While the HD*Calc framework has been incorporated into free software that can be downloaded from NIH websites, the software has two relevant limitations: (1) it is available only on Windows, and (2) it assumes that data is already in a format called a "health status" or "health rate", such as average life expectancy across groups or mortality rates. In our work, we used the HD*Calc framework but created our own software to conduct the calculations on the surrogate features.

There were some limitations to this study. First, note that the work presented here serves the first demonstration of the potential of merging data extracted from medical images into HD index calculations. Due to data accessibility, we limited our investigation to two SDOH demographic correlates, race and sex. Other possibilities, including intersectional identity of race and sex as well as SDOH, will depend upon availability of metadata in datasets. Second, because our goal was to rapidly report a feasibility demonstration study for extracting health disparities data from medical images, we did not incorporate clinical health measures associated with the lungs, such as smoking status. For example, a researcher interested in health disparities in smoking might draw upon separate findings that the likelihood of smoking itself is associated with SDOH,[58] and potentially combine that data with iHDIs from a collection of lung images acquired from known smokers. Third, for iHDIs to be used in HD research, additional extensive work would be needed to identify which medical imaging data source and radiomic features are most practical and useful as well as validate their association with differences in health outcomes. Collaboration among medical imaging scientists, radiologists, informaticians, biostatisticians, and HD researchers will be crucial. Additionally, it would be useful to present the iHDIs with measures of variance, such as 95% confidence intervals, to enhance statistical analysis. However, at the time of manuscript publication, documentation for calculating the variance of the measures was not available for HD*Calc. Lastly, the data used in the study was drawn from one institution, and the SDOH demographic correlates are representative of only this area. Future studies will expand the use of iHDIs to multiple institutions and geographic locations.

In summary, for the first time, the use of medical images as a data source for HD research, via high through-put data extraction and merging for a data input to HD indices, has been demonstrated.

***Conflicts of interest***

M.L.G. is a stockholder in R2 technology/Hologic and shareholder in QView, receives royalties from various companies through the University of Chicago Polsky Center for Entrepreneurship and Innovation, and was a cofounder in Quantitative Insights (now Qlarity Imaging). K.D. and H.L. receive royalties through the University of Chicago Polsky Center for Entrepreneurship and Innovation. It is in the University of Chicago Conflict of Interest Policy that investigators disclose publicly actual or potential significant financial interest that would reasonably appear to be directly and significantly affected by the research activities.

***Funding***

This work was supported by NIMHD P50MD017349-01, NIA K24AG069080-06, and The Medical Imaging Data Resource Center (MIDRC), which is funded by the National Institute of Biomedical Imaging and Bioengineering under Contract No. 75N92020D00021 and by ARPA-H.